\documentclass{elsarticle}
\usepackage{latexsym}
\usepackage{float}
\usepackage[latin2]{inputenc}
\usepackage{graphicx}
\usepackage{ae}
\usepackage{aecompl}
\usepackage{setspace}
\usepackage{amsmath}
\usepackage{amssymb}
\usepackage{amsthm}
\usepackage{booktabs}
\usepackage{longtable}
\def\fn{\hbox{\sl fn\,}}
\def\fl{\hbox{\sl fl\,}}
\def\fa{\hbox{\sl fa\,}}
\def\con{\hbox{\sl con\,}}

\usepackage[hidelinks]{hyperref}

\usepackage{verbatim}

\makeatletter
\def\ps@pprintTitle{%
	\let\@oddhead\@empty
	\let\@evenhead\@empty
	\def\@oddfoot{\centerline{\thepage}}%
	\let\@evenfoot\@oddfoot}
\makeatother

\usepackage[margin=1.2in]{geometry}

\begin{document}

\title{Discovering Sex and Age Implicator Edges in the Human Connectome}
	
\author[p]{László Keresztes\corref{cor2}}
\ead{keresztes@pitgroup.org}
\author[p]{Evelin Szögi\corref{cor2}}
\ead{szogi@pitgroup.org}
\author[p]{Bálint Varga}
\ead{balorkany@pitgroup.org}
\author[p,u]{Vince Grolmusz\corref{cor1}}
\ead{grolmusz@pitgroup.org}
\cortext[cor1]{Corresponding author}
\cortext[cor2]{Joint first authors}
\address[p]{PIT Bioinformatics Group, Eötvös University, H-1117 Budapest, Hungary}
\address[u]{Uratim Ltd., H-1118 Budapest, Hungary}

\date{}

\begin{abstract}
Determining important vertices in large graphs (e.g., Google's PageRank in the case of the graph of the World Wide Web) facilitated the construction of excellent web search engines, returning the most important hits corresponding to the submitted user queries. Interestingly, finding important edges -- instead of vertices -- in large graphs has received much less attention until now. Here we examine the human structural braingraph (or connectome), identified by diffusion magnetic resonance imaging (dMRI) methods, with edges connecting cortical and subcortical gray matter areas and weighted by fiber strengths, measured by the number of the discovered fiber tracts along the edge. We identify several ``single'' important edges in these braingraphs, whose high or low weights imply the sex or the age of the subject observed. We call these edges implicator edges since solely from their weight, one can infer the sex of the subject with more than 67 \% accuracy or their age group with more than 62\% accuracy. We argue that these brain connections are the most important ones characterizing the sex or the age of the subjects. Surprisingly, the edges implying the male sex are mostly located in the anterior parts of the brain, while those implying the female sex are mostly in the posterior regions. Additionally, most of the inter-hemispheric implicator edges are male ones, while the intra-hemispheric ones are predominantly female edges. Our pioneering method for finding the sex- or age implicator edges can also be applied for characterizing other biological and medical properties, including neurodegenerative- and psychiatric diseases besides the sex or the age of the subject, if large and high-quality neuroimaging datasets become available.
We emphasize that our contribution identifies statistically valid single brain connections related to the sex and the age of the subjects in a large and robust dataset. To our knowledge, our results are unprecedented in this aspect. 
\end{abstract}

\maketitle

\bigskip

\section*{Introduction} 

Introducing novel tools in the life sciences has the ultimate goal of finding new pharmaceuticals and therapies for human diseases. In spite of the astronomical increase of the funding of medical research, the yearly number of the globally approved new drugs from the eighties through the first decade of the new century was decreasing on average (\cite{Nicolaou2014}, Figure 1). This observation shows that mankind needs to exploit more aggressively the possibilities, which groundbreaking technologies, methods, and knowledge offered to life sciences. 

One particular challenge is discovering the relations between the fine structure and the function of our brain. For this goal, among other important tools, magnetic resonance imaging (MRI) has a special role: the number of available MRI scanners is constantly increasing worldwide in hospitals and research facilities, and the accompanying data acquisition procedure is non-invasive and harmless. Therefore, the cerebral MRI has the potential application in numerous conditions, assuming we know what to look for in the data. 

A multitude of recent publications examine the human structural connectome, i.e., the macroscopic-scale cerebral connections between the distinct brain areas \cite{Hagmann2012,VanEssen2012,Sporns2005}. These connections were mapped by applying diffusion MRI (dMRI), and using an algorithmic workflow for mapping the braingraph \cite{Daducci2012,Fischl2012,Kerepesi2016b,Kerepesi2015b,Szalkai2016d}. 

Unfortunately, most contributions in this area present findings in the human structural connectome, which cannot be translated to the needs of the medical practice. Some contributions even use weakly defined philosophical terms in the analysis of these graphs instead of strict and clear notations. Our research group has pioneered in applying less philosophical and much more engineering-oriented, exact graph-theoretical terms and definitions, rooted in the classical (mathematical) graph theory, and applied widely in computer engineering \cite{Leighton2014}. For example, we have studied quantitative connectivity-related graph parameters, like the size of the balanced minimum cut, the measure of the graph expanding property, or the size of the minimum vertex cover in \cite{Szalkai2015, Szalkai2016a, Szalkai2017c, Szalkai2016c}, and proved that these parameters are significantly better in women's braingraphs than in men's \cite{Szalkai2015c,Szalkai2015} (where the ``better'' adjective refers to the network connection properties, characterizing better interconnection networks, as defined and studied in \cite{Leighton2014, Dally2007}; it is still not proven that better network parameters directly correspond to better brain functions). We have mapped the frequencies of the edges in consensus braingraphs in several settings \cite{Szalkai2015a,Szalkai2016,Kerepesi2015a}, and discovered frequent subgraphs \cite{Fellner2017}, frequent complete graphs \cite{Fellner2019} and frequent neighbor sets of one of the most widely studied brain areas, the hippocampus \cite{Fellner2018,Fellner2019a}. 

In the medical practice, simpler markers would be much more applicable: for example, the strengths (defined in fiber numbers) of certain, {\em  single} connections between specific brain areas can be measured relatively easily from the diffusion MRIs, and these single edge strengths may imply relevant biological or medical conditions. Similarly, evaluating the consequences of cerebral traumas is easier if we have strong statistical knowledge on the importance of distinct connections between the areas of the gray matter, which are affected by the traumas: that is, not only the importance of the cortical and subcortical gray matter areas need to be evaluated in trauma, but also the importance of the broken connections between these (potentially healthy) gray matter areas.

In the contribution \cite{Keresztes2019} we have found very few edges with dramatic biological effects: we have shown that just 102 edges simply, by a linear relation, determine the sex of the subject, without errors, and, even more surprisingly, we described two superfeminine edges in the human connectome, whose high edge-weights imply the female sex of the subject, independently of the other edges. Similarly, in \cite{Keresztes2019},  we have identified two super-masculine edges with similar properties.

In the present work, we describe several {\em single} human brain connections, whose strengths have strong correlations with the sex or the age of the subjects. For example, the edge connecting the left superior parietal area with the Left-Caudate nucleus has very significantly (p=$10^{-30}$) more axonal tracts in females than in males, and by the observation of only that single edge, one can infer the sex of the subject with 67\% accuracy; more exactly, the following implications are valid with 67\% accuracy:

\medskip

$$\fn( \hbox{ left superiorparietal area, Left-Caudate }) \geq 1561 \Longrightarrow \hbox{ the subject is female;}$$

\medskip

\centerline{ and}

\medskip

$$\fn( \hbox{ left superiorparietal area, Left-Caudate }) < 1561 \Longrightarrow \hbox{ the subject is male.}$$

\medskip

Here $\fn$ denotes the number of observed axonal fibers (i.e., {\bf f}iber {\bf n}umber) in the edge.

 Conversely, the edge connecting the left rostral middle frontal area with the left Caudate nucleus is very significantly stronger in males than in females; one can make sex-identification solely by observing that edge with 67\% accuracy:

\medskip

$$\fn( \hbox{ left rostralmiddlefrontal area, Left-Caudate } ) \geq 1464 \Longrightarrow \hbox{ the subject is male;}$$

\medskip

\centerline{ and}

\medskip
$$\fn( \hbox{ left rostralmiddlefrontal area, Left-Caudate }) < 1464 \Longrightarrow \hbox{ the subject is female.}$$

\medskip

For further implications, we refer to Tables 1 and 2 and the supporting Tables S1, S2, S3, S4, and S5 in the supporting material).

We argue that instead of the sex or the age of the subjects, other medical and biological attributes can also be described this way if the size and the quality of the underlying dataset allow that analysis. Here we were able to find these implications since we have used the high quality 1200 subjects data release of the Human Connectome Project \cite{VanEssen2013a}, and computed the braingraphs with an unprecedentedly robust, error-correcting way \cite{Varga2020}. Our robust source graphs are publicly available at \url{https://braingraph.org} \cite{Varga2020}.

\section*{Methods}

The diffusion MRI data were recorded and made public by the Human Connectome Project's (HCP) website at \url{http://www.humanconnectome.org} \cite{McNab2013} as the 1200-subjects public release. Our group constructed the braingraphs from the HCP data, which are analyzed in the present contribution. The details of the robust graph construction are described in \cite{Varga2020}; here, we just concisely review the graph computation process of \cite{Varga2020} as follows:

\begin{itemize}
	\item[(i)]Parcellation was done by using the Connectome Mapper Tool Kit (CMTK) \cite{Daducci2012} with Lausanne2008 scheme, with labels \url{https://github.com/LTS5/cmp_nipype/blob/master/cmtklib/data/parcellation/lausanne2008/ParcellationLausanne2008.xls}. Five resolutions were computed, with  83, 129, 234, 463 and 1015 nodes. 
	
	\item[(ii)] Probabilistic tractography with random seeding was performed by the MRtrix 0.3 tractography algorithm \cite{Tournier2012}, with 10 repetitions for each subject. A graph edge $\{a,b\}$ connects vertex $a$, corresponding to a gray matter area $A$ and $b$, corresponding to gray matter area $B$ if in each of the 10 runs at least one axonal fiber was found by the tractography algorithm between areas $A$ and $B$. In this case, because of the repeated runs, edge $\{a,b\}$ has 10 fiber numbers, several of them distinct, several of them equal.   
	
	\item[(iii)] As a further error-correcting measure, for each edge, we considered the fiber numbers identified in the ten runs. Next, the minimum and the maximum values of the 10 fiber number values were thrown out as extremes, and the remaining 8 fiber numbers were averaged, and this average was assigned to the edge as its robust, averaged fiber number.
	
\end{itemize} 

We were able to perform the computational steps above for 1064 subjects, each with 5 resolutions, with 83, 129, 234, 463, and 1015 nodes, respectively. The resulting braingraphs can be accessed at the \url{https://braingraph.org/cms/download-pit-group-connectomes/} site. 

In the present work, we consider the coarsest, 83-node resolution of these 1064 robust braingraphs exclusively. The edges at the {\url{https://braingraph.org/cms/download-pit-group-connectomes/}} 
site carry three weights: (i) the number of fibers, denoted by $\fn$; (ii) the mean value of the fiber lengths (in mm) of the edge-defining fibers, denoted by {$\fl$}; the (iii) mean fractional anisotropy $\fa$ of the fibers \cite{Basser2011}. 

Since the fiber number for each edge is an average of the eight of the ten tractography runs, even the fiber number $\fn$ is -- typically -- not an integer. From these weights, here we also introduce an electrical conductivity-related edge weight:
 
$$\con = \frac{\fn\cdot \fa}{\fl^2}.$$

The quantity above is slightly related to electrical conductivity, since it is proportional to the $\hbox{(fiber number)} \times \hbox{(fractional anisotropy)}$ product, where fiber number $\fn$ can be seen as the ``width'' of the connection, and the fractional anisotropy as the ``quality'' of the connection, and it is inversely proportional to the square of the length of the connection. 

The union of the edge-sets of 1064 graphs on 83 nodes has 1950 edges: that is, from the ${83\choose 2}=3403$ vertex-pairs, 1950 form edges in at least one of the 1064 graphs, the remaining vertex-pairs do not form edges (i.e., connections) in the graphs.

\subsection*{Implicator edges}

In what follows, we will find single edges in the braingraph, whose weights imply the sex or the age of the subject, with accuracy greater than 60\%. To our knowledge, no such single implicator edges in the human connectome were identified before the present work. For finding those edges, we will consider the 1950 weighted edges, which appear in at least one subject from 1064, and compare the distribution of their weights between the sexes or the age groups.

We intend to identify for each edge $j$ and for each weighting method a cut-value, denoted by $c$ below, such that the weight of the edge would discriminate between sexes or age groups if they are below or above the cut-value $c$. The list of the edges for which such cut-values exist with strong statistical significance is given in the tables and figures below.

\subsection*{Edge assessment}

For the identification of the sex implicator edges, we consider the two-sample Kolmogorov-Smirnov test \cite{Kolm} with different edge weights. First, we describe the method in detail with abstract edge weights; then, we deal with the specific edge weights.

Let $X$ be the matrix describing the edge weights for each subject: the $1064$ rows of $X$ correspond to the subjects, the $1950$ columns to the edges.  For each edge $j=1,\dotsc, 1950$ we determined the empirical distribution functions for males and females, $F_{male, j}$ and $F_{female, j}$ as follows:
$$ F_{male, j}(t) = \frac{1}{n_m} \sum_{i \, \text{male}} \mathbb{I}(X[i,j] < t)$$
$$ F_{female, j}(t) = \frac{1}{n_f} \sum_{i \, \text{female}} \mathbb{I}(X[i,j] < t)$$

Here, $n_m=489$ is the number of males, and $n_f=575$ is the number of females; $\mathbb{I}$ is the indicator function, its value is 1 if the condition in its argument is satisfied, and 0 otherwise. 

For each edge we computed the $D_j$ value (the greatest absolute difference) of the two empirical distribution functions:

$$D_j = \sup_t |F_{male, j}(t) - F_{female, j}(t)|$$

For each edge $j$, our statistical null hypothesis was that $F_{male, j}$ and  $F_{female, j}$ do not differ, and we rejected the null hypothesis if $D_j$ is high enough (to be specified later). The corresponding $p_j$ p-values were computed by the Python Scipy's stats module \cite{SciPy}.

The $D_j$ statistics are related to the best separation of males and females based on the values of edge $j$. Let $n_f$ be the number of females, $n_m$ be the number of males, and let $ACC$ be the accuracy of best separation. We can assume that in the best separation, females have higher values (the other direction is similar). Let $c$ be the cut-value for the best separation, then:
$$ACC = \frac{n_m F_{male, j}(c) + n_f (1-F_{female,j}(c))}{n_m + n_f}$$

If $n_m = n_f$, then:
$$ACC = \frac{1+F_{male,j}(c)-F_{female,j}(c)}{2} = \frac{1+D_j}{2}$$

Similarly if $n_m \approx n_f$, then:
$$ACC \approx \frac{1+D_j}{2}$$

In the ideal case, when $n_m = n_f$, the edge with the lowest p-value is the best separator. When $n_m \approx n_f$, then edges with lower p-values are presumably better separators. Using the Kolmogorov-Smirnov test with the previous connection to $ACC$, we are searching for better separator edges with higher $D_j$ statistics (and lower $p_j$ values). 

The best separation has a direction, and it enables us to define whether an edge is ``male implicator'' or ``female implicator'':
\begin{itemize}

\item[(i)] An edge $j$ is a female implicator edge with a fixed weight scheme if in the best separation females have higher weight values than the cut-value $c$, or male implicator edge if in the best separation males have higher weight values than the cut-value $c$. We will use this terminology for the implicator edges. 

\end{itemize}

We call an edge significant if we reject the null hypothesis. It is possible that we call an edge significant falsely (type I error), that probability is controlled by the p-value. As we call more and more edges significant (because of low p-values), there is an increasing probability that at least one was falsely labeled as significant; this is the FWER (family-wise error rate). When we had the $p_j$ value for each edge, we can rank them as the edge with the lowest p-value is the most significant. We applied the Holm-Bonferroni correction to control FWER \cite{Holm1979}.

In general, let $H_{0j}, H_{1j}$ $(j = 1,\dotsc,m)$ be $m$ pairs of hypotheses, where $H_{0j}$ is the null hypothesis for edge $j$ (edge $j$ is not significant or there is no difference in distributions) and $H_{1j}$ is the alternative hypothesis for edge $j$ (edge $j$ is significant or there is a difference in distributions). Let $m_0$ be the number of true null hypotheses (number of $j$s, where $H_{0j}$ is true). We can assume, that these are the first $m_0$ hypotheses. Let $A_j$ be the event, that we make a type I error on edge $j$ (we call edge $j$ significant, but $H_{0j}$ is true). The p-values tell that $Pr(A_j) = p_j$. Writing the definition of FWER and using Boole's inequality:
$$FWER = Pr(\bigcup_{j=1}^{m} A_j) = Pr(\bigcup_{j=1}^{m_0} A_j) \leq \sum_{j=1}^{m_0} Pr(A_j) = \sum_{j=1}^{m_0} p_j$$

If we define $\alpha$ and we set $\frac{\alpha}{m}$ on threshold for p-values (reject the null hypothesis if $p_j \leq \frac{\alpha}{m}$), then we control FWER with $\alpha$:

$$FWER \leq \sum_{j=1}^{m_0} p_j \leq m_0 \frac{\alpha}{m} \leq \alpha$$

This Holm-Bonferroni correction works in all cases of dependencies in the tests. This correction is sometimes too strict; there is no edge that $p_j \leq \frac{\alpha}{m}$ with a regular value for $\alpha$ (e.g. $\alpha \leq 0.05$). 

In our study, with $\alpha < 1.95 \times 10^{-5}$, 158 significant edges were found with the $\fn$ edge weight, so the probability that any of these 158 ($\fn$ weighted) edges do not differ between males and females is less than $1.95 \times 10^{-5}$. With $\alpha = 10^{-8} \times m$, where $m=1950$ the number of tests, thus each significant edges must have a p-value lower than $10^{-8}$.

See Table \ref{table1} for the most significant 30 implicator edges based on $\fn$ weight and the supporting Table S1 for the 158 significant edges with the $\fn$ weight. 

For the visualization of the locations of the significant edges in the human brain, we refer to Figures 1 and 2 and supporting Figures S1, S2, S3, and S4 in the supporting material. In the figures, male implicator edges are colored to blue, while female implicator edges to red.

\section*{Discussion and Results}

\subsection*{Significant edges weighted by fiber number}

In Table 1 we present the 30 most significant edges (from the 158 ones in Table S1). The index column refers to the ranking, \textit{Vertex 1} and \textit{Vertex 2} are the corresponding brain areas connected by the edge, \textit{M>F} indicates a male implicator edge, that is, an edge, where the ``male'' implication is valid if the weight of the edge is larger than the $c$ \textit{cut-value}, while \textit{M<F} indicates a female implicator edge, that is, an edge where the ``female'' implication is valid if the weight of the edge is larger than the $c$ \textit{cut-value}. The \textit{p\_ks} and \textit{D\_ks} mean the Kolmogorov-Smirnov p-value and statistics for an edge. \textit{ACC} denotes the accuracy of the best separation, \textit{ACC\_ks} denotes the $\frac{1}{2} (1+ \textit{D\_ks})$ Kolmogorov-Smirnov estimation for best separation. \textit{FWER} in the $k$th row shows the FWER upper bound for the first $k$ edges.

The best possible classification available using only one edge is $67 \%$ accurate, and better than $65 \%$ accuracy available for more than 15 edges: these edges have the highest discriminative power.

\begin{table}[H]
	\centering
	\footnotesize
	\setlength\tabcolsep{1.5pt}
\begin{tabular}{rllcrlrrcl}
\toprule
{} &               Vertex 1 &              Vertex 2  & Male ? Female &     c\ \ &   p\_ks &  D\_ks &   ACC &  ACC\_ks &   FWER \\
\midrule
1  &          lh.superiorparietal &                Left-Caudate &           M<F &  1561 &  1e-30 &  0.36 &  0.67 &    0.68 &  2e-27 \\
2  &      lh.rostralmiddlefrontal &                Left-Caudate &           M>F &  1464 &  8e-29 &  0.35 &  0.67 &    0.67 &  1e-25 \\
3  &   rh.caudalanteriorcingulate &     lh.rostralmiddlefrontal &           M>F &     3 &  4e-28 &  0.34 &  0.67 &    0.67 &  8e-25 \\
4  &                 lh.precuneus &                Left-Caudate &           M<F &   436 &  4e-27 &  0.34 &  0.67 &    0.67 &  9e-24 \\
5  &           lh.parsopercularis &         lh.inferiorparietal &           M<F &     2 &  8e-27 &  0.33 &  0.67 &    0.67 &  2e-23 \\
6  &      rh.rostralmiddlefrontal &               Right-Caudate &           M>F &  1544 &  5e-25 &  0.32 &  0.67 &    0.66 &  9e-22 \\
7  &                 rh.precuneus &               Right-Caudate &           M<F &   335 &  1e-24 &  0.32 &  0.66 &    0.66 &  2e-21 \\
8  &  rh.rostralanteriorcingulate &  rh.caudalanteriorcingulate &           M>F &    18 &  2e-24 &  0.32 &  0.66 &    0.66 &  3e-21 \\
9  &   rh.caudalanteriorcingulate &               Right-Caudate &           M>F &  2462 &  3e-24 &  0.32 &  0.66 &    0.66 &  6e-21 \\
10 &  lh.rostralanteriorcingulate &  lh.caudalanteriorcingulate &           M>F &    58 &  4e-24 &  0.32 &  0.66 &    0.66 &  8e-21 \\
11 &           rh.parsopercularis &         rh.inferiorparietal &           M<F &    20 &  5e-24 &  0.32 &  0.65 &    0.66 &  1e-20 \\
12 &                Right-Putamen &              Right-Amygdala &           M>F &    49 &  6e-22 &  0.30 &  0.65 &    0.65 &  1e-18 \\
13 &          rh.superiortemporal &       rh.transversetemporal &           M>F &    74 &  1e-21 &  0.30 &  0.65 &    0.65 &  3e-18 \\
14 &                Right-Caudate &     lh.rostralmiddlefrontal &           M>F &     5 &  7e-21 &  0.30 &  0.64 &    0.65 &  1e-17 \\
15 &                 rh.precuneus &           Right-Hippocampus &           M<F &  1462 &  2e-20 &  0.29 &  0.64 &    0.65 &  4e-17 \\
16 &                 rh.precuneus &               Right-Putamen &           M<F &   290 &  2e-20 &  0.29 &  0.65 &    0.65 &  4e-17 \\
17 &          rh.superiorparietal &               Right-Caudate &           M<F &   830 &  2e-20 &  0.29 &  0.65 &    0.65 &  4e-17 \\
18 &        lh.posteriorcingulate &                Left-Putamen &           M<F &   658 &  3e-20 &  0.29 &  0.64 &    0.65 &  5e-17 \\
19 &           lh.parsopercularis &            lh.supramarginal &           M<F &   145 &  5e-20 &  0.29 &  0.64 &    0.64 &  9e-17 \\
20 &                 Left-Putamen &               Left-Amygdala &           M>F &    39 &  6e-20 &  0.29 &  0.65 &    0.64 &  1e-16 \\
21 &                    rh.insula &              Right-Amygdala &           M>F &    19 &  6e-19 &  0.28 &  0.64 &    0.64 &  1e-15 \\
22 &                  lh.fusiform &            Left-Hippocampus &           M<F &   310 &  7e-19 &  0.28 &  0.64 &    0.64 &  1e-15 \\
23 &          lh.inferiorparietal &        Left-Thalamus-Proper &           M<F &  1249 &  2e-18 &  0.28 &  0.64 &    0.64 &  3e-15 \\
24 &                   lh.lingual &                Left-Putamen &           M<F &     9 &  6e-18 &  0.27 &  0.63 &    0.64 &  1e-14 \\
25 &  rh.rostralanteriorcingulate &               Right-Caudate &           M>F &   103 &  8e-18 &  0.27 &  0.64 &    0.64 &  1e-14 \\
26 &         Right-Accumbens-area &        Left-Thalamus-Proper &           M>F &     2 &  4e-17 &  0.27 &  0.64 &    0.63 &  7e-14 \\
27 &          lh.inferiorparietal &                Left-Putamen &           M<F &   246 &  1e-16 &  0.26 &  0.63 &    0.63 &  3e-13 \\
28 &           rh.parsopercularis &            rh.supramarginal &           M<F &   205 &  2e-16 &  0.26 &  0.63 &    0.63 &  4e-13 \\
29 &          lh.parstriangularis &     lh.rostralmiddlefrontal &           M>F &   266 &  2e-16 &  0.26 &  0.63 &    0.63 &  4e-13 \\
30 &                  lh.bankssts &            Left-Hippocampus &           M<F &   325 &  2e-16 &  0.26 &  0.64 &    0.63 &  5e-13 \\
\bottomrule
\end{tabular}

	\caption{The 30 most significant sex implicator edges. The index column refers to the ranking, \textit{Vertex 1} and \textit{Vertex 2} are the corresponding brain areas connected by the edge, \textit{M>F} indicates an edge, where the ``male'' implication is valid if the weight of the edge is larger than the \textit{''c'' cut-value}, while \textit{M<F} indicates an edge where the ``female'' implication is valid if the weight of the edge is larger than the \textit{''c'' cut-value}. The \textit{p\_ks} and \textit{D\_ks} mean the Kolmogorov-Smirnov p-value and statistics for the edge. \textit{ACC} denotes the accuracy of the ``male'' or ``female'' implication, \textit{ACC\_ks} denotes the $\frac{1}{2} (1+ \textit{D\_ks})$ Kolmogorov-Smirnov estimation for best separation accuracy. \textit{FWER} in the $k$th row shows the FWER upper bound for the first $k$ edges.}
	\label{table1}
\end{table}

We review some of the most significant sex implicator edges from Table 1 here. The higher fiber numbers of the edges between the caudate and the superior parietal regions imply the female sex in both hemispheres (Table 1, lines 1 and 17), with 67\% and 65\% accuracy, respectively.

In contrast, the higher fiber numbers in the edges between the caudate and rostral middle frontal regions in both hemispheres (Table 1, lines 2 and 6) imply the male sex with 67\% accuracy. 

In the literature, the endpoints of the edges in Table 1 frequently appeared in volumetric studies in conjunction with sex dimorphisms and sexual functions. It is not surprising that the caudate nucleus and the putamen appear several times in Table 1 as endpoints of implicator edges: the basal ganglia have a large number of sex steroid receptors, and they were demonstrated having sex dimorphisms in previous volumetric studies \cite{Rijpkema2012, Witte2010}. Other volumetric studies have shown the sex difference in the development in the hippocampus vs. the amygdala: the amygdala volumes increase more in men and the hippocampus volumes more in women, during puberty \cite{Giedd1997}. Volumetric sex differences were also reported in the inferior parietal lobule in \cite{Frederikse1999a}. 

We emphasize that in our contribution, we identify statistically valid {\em brain connections} (and not just the volumes of the brain areas which were previously done by many sources) related to sex and age in a large, robust and coherent dataset.

\subsection*{Plots of the significant edges weighted by fiber numbers ($\fn$)}

In this section, we show the resulting plots of the significant edge selection method. 

The blue edges are always the male implicator edges, and the red edges are the female implicator ones. This categorization is based on the best separation on the corresponding edge as we described.

\begin{figure}[H]
	\centering
	\includegraphics[scale=0.30]{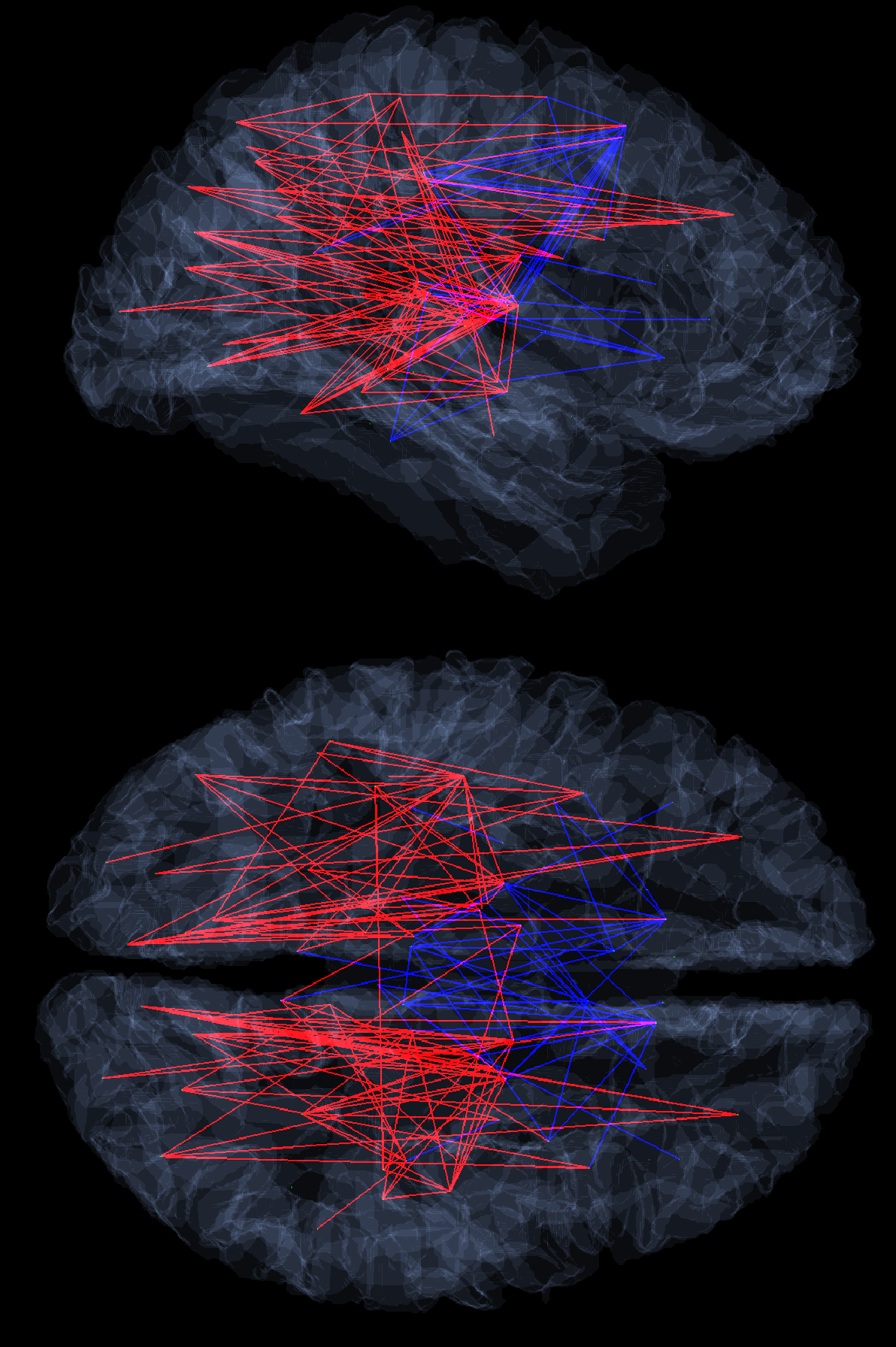}
	\caption{Location of the 158 significant sex-implicator edges with the fiber number ($\fn$) weighting in sagittal and horizontal positions. In red edges, the fiber number is significantly larger in women (they are the female implicator edges). In blue edges, the fiber number is significantly larger in men (they are the male implicator edges). The upper panel shows the brain in the sagittal, the lower panel in the horizontal position. It is very surprising to notice that the blue male implicator edges are more frequent in the anterior parts of the brain, while the red female implicator edges in the posterior lobes. }
	\label{fn1}
\end{figure}

Here we consider the most natural $\fn$ (fiber-number) weight. Figures with other weight functions can be found in the supporting material. We selected $\alpha = 1.95 \times 10^{-5}$ as the upper bound on FWER, which is the same as calling an edge significant if the p-value is lower than $10^{-8}$. With this setting on $\alpha$, 158 significant edges were found (see Table S1). Figure \ref{fn1} shows their location.
 
It is striking that the male implicator edges are located mostly in the anterior, while the female edges in the posterior areas of the brain. 

\begin{figure}[H]
	\centering
	\includegraphics[scale=0.42]{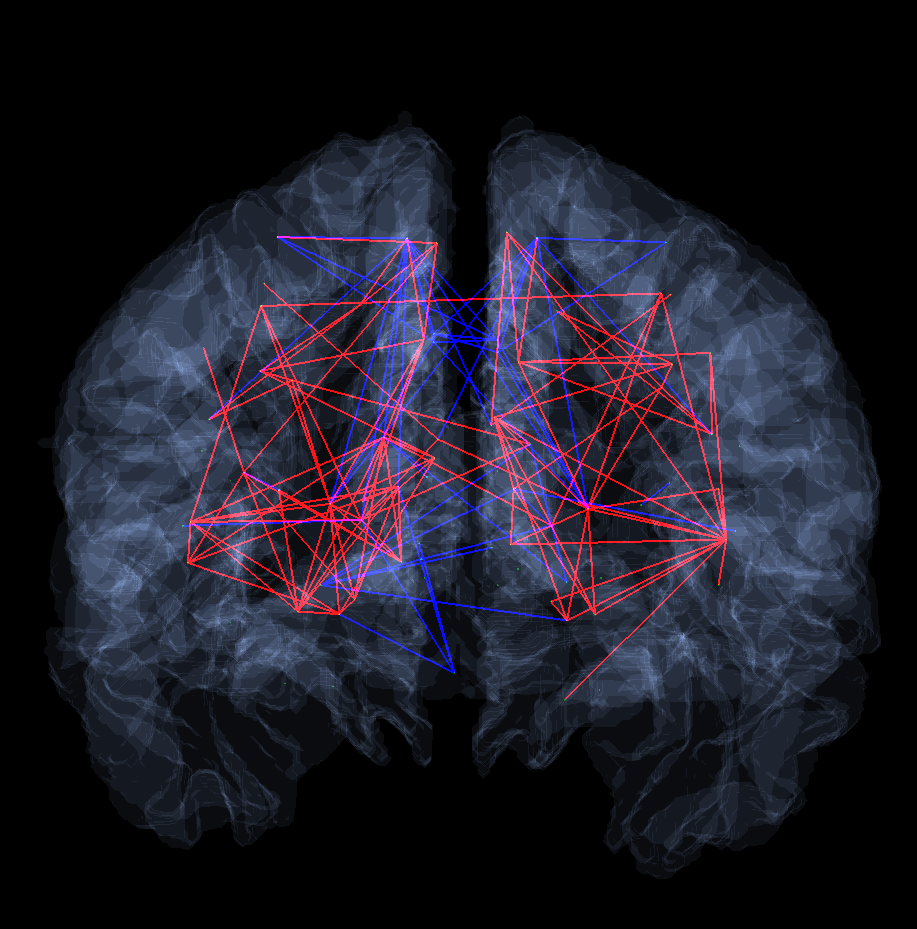}
	\caption{Location of the 158 significant sex-implicator edges with the fiber number ($\fn$) weighting in coronal view. Red edges are the female implicator edges; blue edges are the male implicator edges. One should observe that the inter-hemispheric implicator edges are mostly blue, while the red, female implicator edges are mostly intra-hemispheric ones. }
\end{figure}

For the first view, the findings of Figure 2 are in contrast to the results of \cite{Ingalhalikar2014b}, which stated that the inter-hemispheric connections of males are weaker than females. However, our findings here do not necessarily contradict with \cite{Ingalhalikar2014b}: we do not count the inter-hemispheric and intra-hemispheric connections here, but rather have identified the sex-implicator connections with different weight functions; that is, our method, focus, and analysis are quite distinct from that of \cite{Ingalhalikar2014b}.

\subsection*{Results with weight functions con, fa, and fl}

The tables in the on-line supporting material (Tables S3, S4, and S5) list the edges, implying the male or female sex with cut-values determined in $\con$, $\fa$ and $\fl$ weight functions, respectively. The supporting Figures S2, S3, and S4 visualize the significant sex implicator edges, using these weight functions. Additionally, supporting animations are available for visualizing the implicator edges with fiber number ($\fn$) weight \url{https://youtu.be/VHVPghr8Ms4}, with fiber length ($\fl$) weight \url{https://youtu.be/KVzP5OpxU2E}, with fractional anisotropy ($\fa$) weight \url{https://youtu.be/-bkB2qqXawY }, and conductivity ($\con$) weight \url{https://youtu.be/oQFO3QLuCNY}. 

Table S5 and Figure S4 with the fiber-length weight function have some surprising edges: anatomically, the brains of men are larger than the brains of women by around 7\% in volume \cite{Witelson2006}; therefore, it is natural that most sex implicator edges for males are corresponding to longer brain connections. In other words, in Figure S4, most implicator edges are colored blue. That is, they imply male subjects if longer than a cut-value.

However, some edges are red, meaning that even in the fiber length ($\fl$) weight function, their longer lengths imply female subjects. 

The edges connecting the parsopercularis and precentral areas in both left and right hemispheres are implying the female sex if they are longer, as well as the edge between the putamen and hippocampus, both in the right and left hemispheres (Table S5). These findings are in line with the results of \cite{Goddings2014}, where differences in developmental trajectories of sexes were observed in these areas. Here we first proved that the length of these edges has very strong and implicative sex differences with longer lengths in females.       

We note that our quantitative findings with cut-values and accuracy measures make our method applicable in diffusion MRI-based diagnostic tools for neurodegenerative- and psychiatric diseases if high-quality and large MRI datasets would become available for the public.

\section*{Age implicator edges}

We attempted to apply the previous method on the property of age instead of sex. Probably since the age span of the subjects is not large enough (from 1064 subjects, 224 were between 21 and 25 years; 466 between 26 and 30 years; 364 between 31 and 36; and 10 were older than 36), we achieved only a moderate success.

We were interested in the difference between the 21-25 age group (``young'') and the 31-35 age group (``old''), so we also selected 224 subjects from the ``old'' group and compared them with the ``young'' subjects.

Because of the small number of examples and probably less difference between classes, controlling FWER was not possible, so we have controlled the false discovery rate (FDR) instead of FWER. Given $m$ number of hypotheses, if the number of discoveries (rejections of hypotheses) is $R$, and the number of false discoveries is $V$, then FDR = $E(\frac{V}{R})$. Thus an upper bound FDR $< \alpha$ would tell that the expected rate of false discoveries does not exceed $\alpha$.

In bioinformatics, an upper bound on FDR is still valuable because if the number of rejections is small, then checking these hypotheses experimentally probably reveals the feature of interest (e.g., in gene expression data).

Benjamini and Yekutieli provided a general procedure for controlling FDR with $\alpha$ in cases of arbitrary dependencies \cite{Benjamini2001}. The recipe is as follows: (i) Sort the p-values: $p_1 \leq \dotsc \leq p_m$; (ii) Find the largest $k$ for $p_k \leq \frac{k}{m c(m)} \alpha$, where $c(m) = \sum_{i=1}^m \frac{1}{i}$. (iii) Reject the first $k$ hypotheses.

Table \ref{age1} shows the results.

\begin{table}[H]
	\centering
	\footnotesize
	\setlength\tabcolsep{1.5pt}

\begin{tabular}{lllcrlrccr}
\toprule
{} &              Vertex 1 &         Vertex 2 & Young ? Old &     c &   p\_ks &  D\_ks &   ACC &  ACC\_ks &   FDR \\
\midrule
1  &              rh.paracentral &         Right-Pallidum &         Y>O &    41 &  4e-06 &  0.24 &  0.62 &    0.62 &  0.06 \\
2  &          lh.parsopercularis &    lh.inferiorparietal &         Y<O &     4 &  3e-05 &  0.22 &  0.61 &    0.61 &  0.21 \\
3  &              rh.paracentral &  Right-Thalamus-Proper &         Y>O &    94 &  6e-05 &  0.21 &  0.61 &    0.61 &  0.34 \\
4  &              rh.paracentral &          Right-Putamen &         Y>O &   306 &  6e-05 &  0.21 &  0.61 &    0.61 &  0.26 \\
5  &                   lh.insula &          Left-Pallidum &         Y>O &   230 &  6e-05 &  0.21 &  0.61 &    0.61 &  0.20 \\
6  &  lh.caudalanteriorcingulate &           Left-Caudate &         Y>O &  2903 &  6e-05 &  0.21 &  0.61 &    0.61 &  0.17 \\
7  &      rh.caudalmiddlefrontal &  Right-Thalamus-Proper &         Y>O &   552 &  2e-04 &  0.21 &  0.60 &    0.60 &  0.34 \\
8  &              rh.postcentral &         Right-Pallidum &         Y>O &    37 &  2e-04 &  0.21 &  0.60 &    0.60 &  0.30 \\
9  &  rh.caudalanteriorcingulate &              rh.insula &         Y>O &   144 &  2e-04 &  0.20 &  0.60 &    0.60 &  0.40 \\
10 &               rh.precentral &           rh.precuneus &         Y<O &    54 &  2e-04 &  0.20 &  0.60 &    0.60 &  0.36 \\
\bottomrule
\end{tabular}

	\caption{The ten most significant edges with the $\fn$ weight function in age. The index column refers to the ranking, \textit{Vertex 1} and \textit{Vertex 2} are the corresponding brain areas connected by the edge, \textit{Y>O} indicates an edge where the ``young'' implication is valid if the weight of the edge is larger than the \textit{``c'' cut-value}, while \textit{Y<O} indicates an edge where the ``old'' implication is valid if the weight of the edge is larger than the \textit{``c'' cut-value}. The \textit{p\_ks} and \textit{D\_ks} mean the Kolmogorov-Smirnov p-value and statistics for an edge. \textit{ACC} denotes the accuracy of the implication, \textit{ACC\_ks} denotes the $\frac{1}{2} (1+ \textit{D\_ks})$ Kolmogorov-Smirnov estimation for best separation accuracy. \textit{FDR} is the Benjamini-Yekutieli upper bound for the false discovery rate. Table S2 in the supporting material contains the whole result-set.}
	\label{age1}
\end{table}

For example, if we set $\alpha = 0.17$, then the first 6 edges could be called significant with the assumption that the number of false discoveries would be $\leq 1.02$. Here the \textit{ACC} and \textit{ACC\_ks} are the same; each class contains the same number ($=224$) of elements.

\section*{Conclusions}

We have demonstrated that certain {\em single} brain connections can be applied for inferring biological properties of the subjects with strict statistical validity. We have shown that the fiber number, the fiber length, the fractional anisotropy, and a connectivity-related edge-weight of specific edges can be applied for predicting the sex and the age of the subjects with 67\% and 62\% accuracy, respectively. To our knowledge, this is the first result, which uses the properties of single braingraph edges for gaining statistically valid biological implications. Our method is believed to be applicable in other implications besides age or sex if high-quality diffusion MRI datasets become available.

Apart from the possible diagnostic use, the implicator edges are the most important ones in distinguishing the biological properties examined. This way, we have demonstrated the most important single connections related to specific biological properties in large graphs.

\section*{Author contributions:}  LK and ES suggested using the statistical methods for the identification of single edges with the implications described in this work, contributed statistical methods and analyzed the results. BV computed the braingraphs from the HCP public data and prepared the figures and the supplementary videos accompanying the present work. VG initiated the study, secured funding, analyzed the results, and wrote the paper. 

\section*{Data availability} The data source of this work was published at the Human Connectome Project's website at \url{http://www.humanconnectome.org} \cite{McNab2013} as the 1200-subjects public release. The parcellation data, containing the anatomically labeled ROIs, is listed in the CMTK nypipe GitHub repository \url{https://github.com/LTS5/cmp_nipype/blob/master/cmtklib/data/parcellation/lausanne2008/ParcellationLausanne2008.xls}.  The underlying braingraph set is published and downloadable at \url{https://braingraph.org}, with the methodology described in detail in \cite{Varga2020}. The supporting animations for visualizing the implicator edges with fiber number ($\fn$) weight is available at \url{https://youtu.be/VHVPghr8Ms4}, with fiber length ($\fl$) weight at \url{https://youtu.be/KVzP5OpxU2E}, with fractional anisotropy ($\fa$) weight at \url{https://youtu.be/-bkB2qqXawY }, and conductivity ($\con$) weight at \url{https://youtu.be/oQFO3QLuCNY}.

\section*{Acknowledgments}
Data were provided in part by the Human Connectome Project, WU-Minn Consortium (Principal Investigators: David Van Essen and Kamil Ugurbil; 1U54MH091657) funded by the 16 NIH Institutes and Centers that support the NIH Blueprint for Neuroscience Research; and by the McDonnell Center for Systems Neuroscience at Washington University. VG and BV were partially supported by the VEKOP-2.3.2-16-2017-00014 program, supported by the European Union and the State of Hungary, co-financed by the European Regional Development Fund, VG by NKFI-127909
grants of the National Research, Development and Innovation Office of Hungary. LK and ES were supported in part by the EFOP-3.6.3-VEKOP-16-2017-00002 grant, supported by the European Union, co-financed by the European Social Fund.

\bigskip 

%\bibliography{v:/vince/CIKKEK/medl}
%\bibliographystyle{unsrt}

\section*{Supporting Material}

\section*{Results with the fiber number (fn) weight function}

\begin{figure}[H]
	\centering
	\includegraphics[width=16cm]{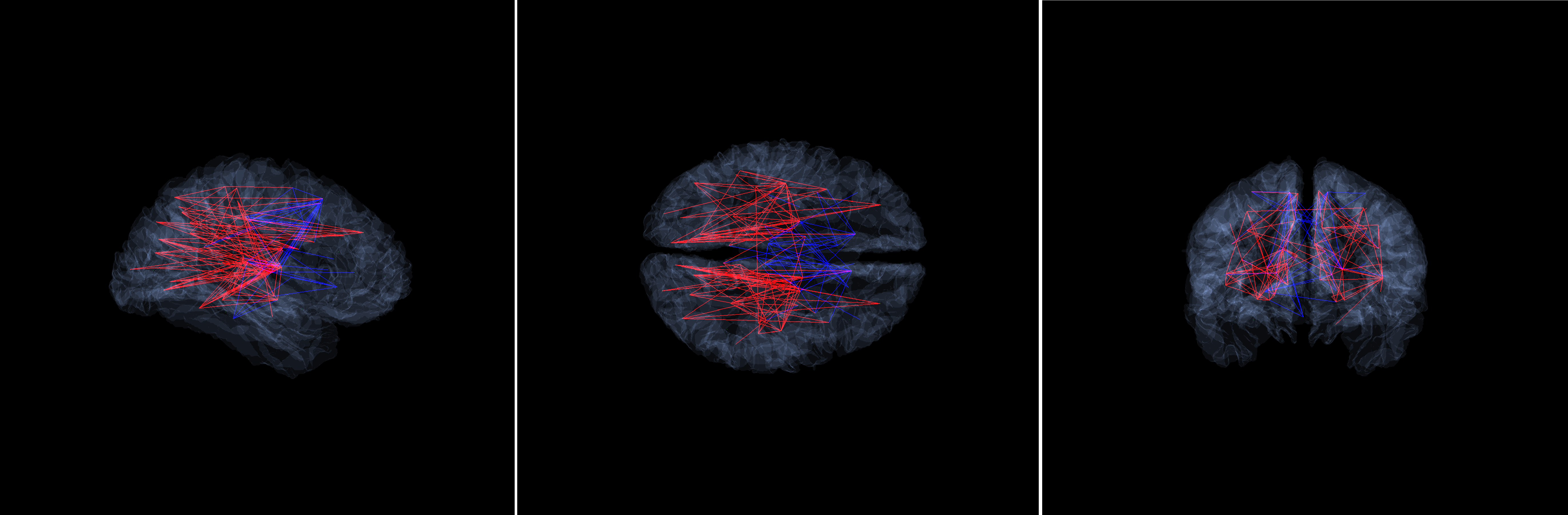}
\end{figure}
\noindent {\bf Figure S1.} {\sl Sex implicator edges with the fiber number (fn) weight function. Red color denotes edges where the higher fiber number implies the female sex, while blue edges correspond to those where the higher fiber number implies the male sex (cf. with Table S1). Three views correspond (from left to right) to sagittal, horizontal, and coronal positions. One may observe that red edges are mostly located in the posterior, while blue edges in the anterior lobes. Additionally, the inter-hemispheric edges are mostly blue ones. An animation of these images is available at \url{https://youtu.be/VHVPghr8Ms4}}. 

\section*{Table S1: Sex implicator edges with the fiber number (fn) weight function}

The row index shows the ranking, \textit{Vertex 1} and \textit{Vertex 2} are the corresponding brain areas, connected by the edge, \textit{M>F} indicates an edge with weight function significantly higher in males, while \textit{M<F} indicates an edge with weight function significantly higher in females. The \textit{p\_ks} and \textit{D\_ks} mean the Kolmogorov-Smirnov p-value and statistics for an edge. \textit{ACC} denotes the accuracy of the best separation, \textit{ACC\_ks} denotes the $\frac{1}{2} (1+ \textit{D\_ks})$ Kolmogorov-Smirnov estimation for best separation accuracy. \textit{FWER} in the $k$th row shows the Family-Wise Error Rate upper bound for the first $k$ edges.

{\small

\begin{longtable}{rllclrrrl}

\toprule
{} &               Vertex 1 &               Vertex 2 & Male ? Female &   p\_ks &  D\_ks &   ACC &  ACC\_ks &   FWER \\
\midrule
1   &          lh.superiorparietal &                 Left-Caudate &           M<F &  1e-30 &  0.36 &  0.67 &    0.68 &  2e-27 \\
2   &      lh.rostralmiddlefrontal &                 Left-Caudate &           M>F &  8e-29 &  0.35 &  0.67 &    0.67 &  1e-25 \\
3   &   rh.caudalanteriorcingulate &      lh.rostralmiddlefrontal &           M>F &  4e-28 &  0.34 &  0.67 &    0.67 &  8e-25 \\
4   &                 lh.precuneus &                 Left-Caudate &           M<F &  4e-27 &  0.34 &  0.67 &    0.67 &  9e-24 \\
5   &           lh.parsopercularis &          lh.inferiorparietal &           M<F &  8e-27 &  0.33 &  0.67 &    0.67 &  2e-23 \\
6   &      rh.rostralmiddlefrontal &                Right-Caudate &           M>F &  5e-25 &  0.32 &  0.67 &    0.66 &  9e-22 \\
7   &                 rh.precuneus &                Right-Caudate &           M<F &  1e-24 &  0.32 &  0.66 &    0.66 &  2e-21 \\
\bottomrule

\end{longtable}

}

\section*{Table S2: Age implicator edges with the fiber number (fn) weight function}

The row index shows the ranking, \textit{Vertex 1} and \textit{Vertex 2} are the corresponding brain areas, \textit{Y>O} indicates an edge where the ``young'' implication is valid if the weight of the edge is larger than the \textit{``c'' cut-value}, while \textit{Y<O} indicates an edge where the ``old'' implication is valid if the weight of the edge is larger than the \textit{``c'' cut value}. The \textit{p\_ks} and \textit{D\_ks} mean the Kolmogorov-Smirnov p-value and statistics for an edge. \textit{ACC} denotes the accuracy of the best separation, \textit{ACC\_ks} denotes the $\frac{1}{2} (1+ \textit{D\_ks})$ Kolmogorov-Smirnov estimation for best separation accuracy. \textit{FDR} is the Benjamini-Yekutieli upper bound to the false discovery rate.
{\small
    \begin{longtable}{llllrlrrrr}
\toprule
{} &              Vertex 1 &               Vertex 2 & Young ? Old &     c &   p\_ks &  D\_ks &   ACC &  ACC\_ks &   FDR \\
\midrule
1  &              rh.paracentral &               Right-Pallidum &         Y>O &    41 &  4e-06 &  0.24 &  0.62 &    0.62 &  0.06 \\
2  &          lh.parsopercularis &          lh.inferiorparietal &         Y<O &     4 &  3e-05 &  0.22 &  0.61 &    0.61 &  0.21 \\
3  &              rh.paracentral &        Right-Thalamus-Proper &         Y>O &    94 &  6e-05 &  0.21 &  0.61 &    0.61 &  0.34 \\
4  &              rh.paracentral &                Right-Putamen &         Y>O &   306 &  6e-05 &  0.21 &  0.61 &    0.61 &  0.26 \\
5  &                   lh.insula &                Left-Pallidum &         Y>O &   230 &  6e-05 &  0.21 &  0.61 &    0.61 &  0.20 \\
6  &  lh.caudalanteriorcingulate &                 Left-Caudate &         Y>O &  2903 &  6e-05 &  0.21 &  0.61 &    0.61 &  0.17 \\
7  &      rh.caudalmiddlefrontal &        Right-Thalamus-Proper &         Y>O &   552 &  2e-04 &  0.21 &  0.60 &    0.60 &  0.34 \\
8  &              rh.postcentral &               Right-Pallidum &         Y>O &    37 &  2e-04 &  0.21 &  0.60 &    0.60 &  0.30 \\
9  &  rh.caudalanteriorcingulate &                    rh.insula &         Y>O &   144 &  2e-04 &  0.20 &  0.60 &    0.60 &  0.40 \\
10 &               rh.precentral &                 rh.precuneus &         Y<O &    54 &  2e-04 &  0.20 &  0.60 &    0.60 &  0.36 \\
11 &          lh.superiorfrontal &         Left-Thalamus-Proper &         Y>O &   147 &  3e-04 &  0.20 &  0.60 &    0.60 &  0.49 \\
12 &     rh.rostralmiddlefrontal &                Right-Caudate &         Y>O &  1597 &  5e-04 &  0.19 &  0.60 &    0.60 &  0.67 \\
13 &               rh.precentral &               Right-Pallidum &         Y>O &    41 &  7e-04 &  0.19 &  0.59 &    0.59 &  0.90 \\
14 &       rh.posteriorcingulate &             rh.supramarginal &         Y>O &   426 &  7e-04 &  0.19 &  0.59 &    0.59 &  0.84 \\
15 &     lh.rostralmiddlefrontal &  lh.rostralanteriorcingulate &         Y>O &    72 &  7e-04 &  0.19 &  0.59 &    0.59 &  0.78 \\
16 &               rh.precentral &        Right-Thalamus-Proper &         Y>O &   387 &  7e-04 &  0.19 &  0.59 &    0.59 &  0.73 \\
17 &      rh.caudalmiddlefrontal &               Right-Pallidum &         Y>O &    77 &  7e-04 &  0.19 &  0.59 &    0.59 &  0.69 \\
18 &               rh.precentral &             rh.supramarginal &         Y<O &  2008 &  1e-03 &  0.18 &  0.59 &    0.59 &  0.95 \\
19 &               lh.precentral &         Left-Thalamus-Proper &         Y>O &   567 &  1e-03 &  0.18 &  0.59 &    0.59 &  0.90 \\
20 &            rh.supramarginal &          rh.superiortemporal &         Y<O &    40 &  1e-03 &  0.18 &  0.59 &    0.59 &  0.85 \\
\bottomrule
 \end{longtable}%
  \label{tab:addlabel}%
}

\section*{The conductivity (con) weight function}

\begin{figure}[H]
	\centering
	\includegraphics[width=16cm]{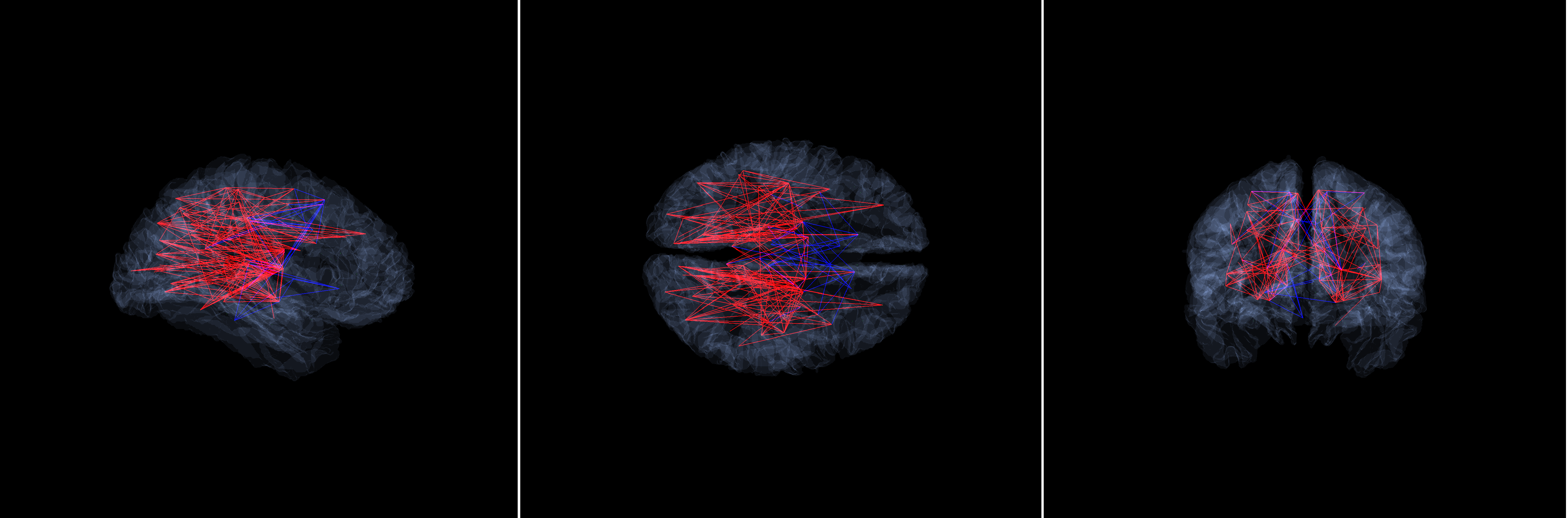}
\end{figure}
\noindent {\bf Figure S2.} {\sl 193 sex implicator edges with the conductivity (con) weight function with FWER $< 1.95 \times 10^{-5}$ . Red color denotes edges where the higher con weight implies the female sex, while blue edges correspond to those where higher con weight implies the male sex (cf.  Table S3). Three views correspond (from left to right) to sagittal, horizontal, and coronal positions. One may observe that red edges are mostly located in the posterior, while blue edges in the anterior lobes. Additionally, the inter-hemispheric edges are mostly blue ones. An animation of these images is available at \url{https://youtu.be/oQFO3QLuCNY}}.

\section*{Table S3: Sex implicator edges with the conductivity (con) weight function}

193 edges with FWER $< 1.95 \times 10^{-5}$ with the conductivity weight function. Index shows the ranking, \textit{Vertex 1} and \textit{Vertex 2} are the corresponding brain areas, \textit{M>F} indicates an edge with weight function significantly higher in males, while \textit{M<F} indicates an edge with weight function significantly higher in females. The \textit{p\_ks} and \textit{D\_ks} mean the Kolmogorov-Smirnov p-value and statistics for an edge. \textit{ACC} denotes the accuracy of the best separation, \textit{ACC\_ks} denotes the $\frac{1}{2} (1+ \textit{D\_ks})$ KS estimation for best separation accuracy. \textit{FWER} in the $k$th row shows the FWER upper bound for the first $k$ edges.

{\small
\begin{longtable}{rllclrrrl}

\toprule
{} &               Vertex 1 &               Vertex 2 & Male ? Female &   p\_ks &  D\_ks &   ACC &  ACC\_ks &   FWER \\
\midrule
1   &                 rh.precuneus &                Right-Putamen &           M<F &  4e-30 &  0.35 &  0.68 &    0.68 &  8e-27 \\
2   &        Right-Thalamus-Proper &         Left-Thalamus-Proper &           M<F &  4e-30 &  0.35 &  0.68 &    0.68 &  8e-27 \\
3   &          lh.inferiorparietal &         Left-Thalamus-Proper &           M<F &  1e-28 &  0.35 &  0.67 &    0.67 &  2e-25 \\
4   &                 lh.precuneus &                 Left-Caudate &           M<F &  6e-28 &  0.34 &  0.66 &    0.67 &  1e-24 \\
5   &           lh.parsopercularis &          lh.inferiorparietal &           M<F &  3e-27 &  0.34 &  0.67 &    0.67 &  6e-24 \\
6   &                 rh.precuneus &                Right-Caudate &           M<F &  1e-26 &  0.33 &  0.67 &    0.67 &  2e-23 \\
7   &          lh.superiorparietal &                 Left-Caudate &           M<F &  1e-26 &  0.33 &  0.67 &    0.67 &  2e-23 \\
\bottomrule

\end{longtable}
}

\section*{The fractional anisotropy (fa) weight function}

\begin{figure}[H]
	\centering
	\includegraphics[width=16cm]{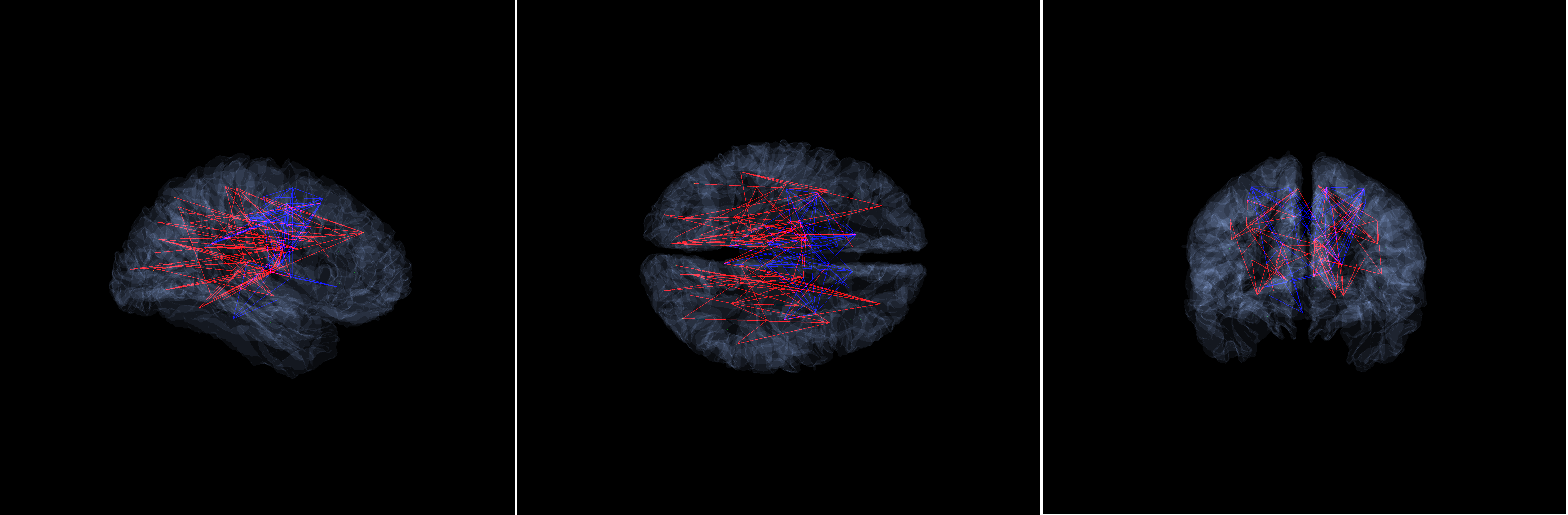}
\end{figure}
\noindent {\bf Figure S3.} {\sl 120 sex implicator edges with the fractional anisotropy (fa) weight function with FWER < $1.95 \times 10^{-5}$. Red color denotes edges where the higher fa weight implies the female sex, while blue edges correspond to those where higher fa weight implies the male sex (cf. Table S4). Three views correspond (from left to right) to sagittal, horizontal, and coronal positions. One may observe that red edges are mostly located in the posterior, while blue edges in the anterior lobes. Additionally, the inter-hemispheric edges are mostly blue ones. An animation of these images is available at \url{https://youtu.be/-bkB2qqXawY}}.

\section*{Table S4: Sex implicator edges with the fractional anisotropy (fa) weight function}

120 sex implicator edges with the fractional anisotropy (fa) weight function with FWER < $1.95 \times 10^{-5}$. Index shows the ranking, \textit{Vertex 1} and \textit{Vertex 2} are the corresponding brain areas, \textit{M>F} indicates an edge with weight function significantly higher in males, while \textit{M<F} indicates an edge with weight function significantly higher in females. The \textit{p\_ks} and \textit{D\_ks} mean the Kolmogorov-Smirnov p-value and statistics for an edge. \textit{ACC} denotes the accuracy of the best separation, \textit{ACC\_ks} denotes the $\frac{1}{2} (1+ \textit{D\_ks})$ KS estimation for best separation accuracy. \textit{FWER} in the $k$th row shows the FWER upper bound for the first $k$ edges.

{\small
\begin{longtable}{rllclrrrl}
\toprule
{} &               Vertex 1 &               Vertex 2 & Male ? Female &   p\_ks &  D\_ks &   ACC &  ACC\_ks &   FWER \\
\midrule
1   &   rh.caudalanteriorcingulate &      lh.rostralmiddlefrontal &           M>F &  5e-25 &  0.32 &  0.66 &    0.66 &  1e-21 \\
2   &           lh.parsopercularis &          lh.inferiorparietal &           M<F &  2e-24 &  0.32 &  0.66 &    0.66 &  4e-21 \\
3   &           lh.parsopercularis &       lh.caudalmiddlefrontal &           M<F &  5e-23 &  0.31 &  0.65 &    0.65 &  1e-19 \\
4   &      rh.rostralmiddlefrontal &       rh.caudalmiddlefrontal &           M>F &  6e-23 &  0.31 &  0.67 &    0.65 &  1e-19 \\
5   &           rh.superiorfrontal &        Right-Thalamus-Proper &           M>F &  1e-22 &  0.31 &  0.65 &    0.65 &  2e-19 \\
6   &           rh.superiorfrontal &       rh.caudalmiddlefrontal &           M>F &  5e-22 &  0.30 &  0.66 &    0.65 &  1e-18 \\
7   &          rh.parstriangularis &                rh.precentral &           M<F &  1e-21 &  0.30 &  0.65 &    0.65 &  2e-18 \\
8   &      rh.rostralmiddlefrontal &           rh.superiorfrontal &           M>F &  3e-21 &  0.30 &  0.65 &    0.65 &  5e-18 \\
9   &      rh.rostralmiddlefrontal &   rh.caudalanteriorcingulate &           M>F &  9e-21 &  0.29 &  0.65 &    0.65 &  2e-17 \\
10  &                Right-Putamen &               Right-Amygdala &           M>F &  2e-19 &  0.29 &  0.63 &    0.64 &  3e-16 \\
11  &          lh.parstriangularis &                lh.precentral &           M<F &  3e-19 &  0.28 &  0.63 &    0.64 &  7e-16 \\
12  &           rh.parsopercularis &                rh.precentral &           M<F &  4e-19 &  0.28 &  0.65 &    0.64 &  8e-16 \\
13  &                    rh.insula &               Right-Amygdala &           M>F &  2e-18 &  0.28 &  0.63 &    0.64 &  4e-15 \\
14  &  rh.rostralanteriorcingulate &   rh.caudalanteriorcingulate &           M>F &  3e-18 &  0.28 &  0.63 &    0.64 &  5e-15 \\
15  &      rh.rostralmiddlefrontal &   lh.caudalanteriorcingulate &           M>F &  3e-18 &  0.28 &  0.64 &    0.64 &  5e-15 \\
16  &      rh.rostralmiddlefrontal &        rh.posteriorcingulate &           M>F &  4e-18 &  0.27 &  0.64 &    0.64 &  8e-15 \\
17  &      rh.rostralmiddlefrontal &                Right-Caudate &           M>F &  2e-17 &  0.27 &  0.63 &    0.63 &  4e-14 \\
18  &         Right-Accumbens-area &         Left-Thalamus-Proper &           M>F &  2e-17 &  0.27 &  0.64 &    0.63 &  4e-14 \\
19  &          rh.parstriangularis &               rh.postcentral &           M<F &  3e-17 &  0.27 &  0.63 &    0.63 &  6e-14 \\
20  &           rh.superiorfrontal &   rh.caudalanteriorcingulate &           M>F &  4e-17 &  0.27 &  0.63 &    0.63 &  8e-14 \\
21  &                   rh.lingual &                    rh.insula &           M<F &  1e-16 &  0.26 &  0.62 &    0.63 &  2e-13 \\
22  &           lh.superiorfrontal &         Left-Thalamus-Proper &           M>F &  3e-16 &  0.26 &  0.62 &    0.63 &  6e-13 \\
23  &           rh.superiorfrontal &                Right-Caudate &           M>F &  3e-15 &  0.25 &  0.62 &    0.63 &  7e-12 \\
24  &      rh.rostralmiddlefrontal &                 Left-Caudate &           M>F &  3e-15 &  0.25 &  0.63 &    0.63 &  7e-12 \\
25  &        Right-Thalamus-Proper &                Right-Putamen &           M<F &  3e-15 &  0.26 &  0.63 &    0.63 &  7e-12 \\
26  &      rh.rostralmiddlefrontal &        Right-Thalamus-Proper &           M>F &  3e-15 &  0.26 &  0.64 &    0.63 &  7e-12 \\
27  &          rh.parstriangularis &             rh.supramarginal &           M<F &  4e-15 &  0.25 &  0.62 &    0.62 &  9e-12 \\
28  &           lh.parsopercularis &                lh.precentral &           M<F &  8e-15 &  0.25 &  0.62 &    0.62 &  2e-11 \\
29  &                   lh.lingual &                 Left-Caudate &           M<F &  1e-14 &  0.25 &  0.62 &    0.62 &  2e-11 \\
30  &                 rh.precuneus &                Right-Putamen &           M<F &  1e-14 &  0.25 &  0.63 &    0.62 &  2e-11 \\
31  &           rh.superiorfrontal &               Right-Pallidum &           M>F &  1e-14 &  0.25 &  0.63 &    0.62 &  2e-11 \\
32  &                Right-Caudate &  lh.rostralanteriorcingulate &           M>F &  1e-14 &  0.25 &  0.63 &    0.62 &  2e-11 \\
33  &           lh.superiorfrontal &                Left-Pallidum &           M>F &  2e-14 &  0.25 &  0.62 &    0.62 &  3e-11 \\
34  &      rh.rostralmiddlefrontal &        lh.posteriorcingulate &           M>F &  3e-14 &  0.24 &  0.63 &    0.62 &  7e-11 \\
35  &                rh.precentral &               rh.postcentral &           M<F &  4e-14 &  0.24 &  0.63 &    0.62 &  7e-11 \\
36  &                 Left-Putamen &                Left-Amygdala &           M>F &  6e-14 &  0.24 &  0.61 &    0.62 &  1e-10 \\
37  &               lh.postcentral &          lh.superiortemporal &           M<F &  6e-14 &  0.24 &  0.61 &    0.62 &  1e-10 \\
38  &                Left-Pallidum &                Left-Amygdala &           M>F &  8e-14 &  0.24 &  0.61 &    0.62 &  2e-10 \\
39  &                   lh.lingual &                    lh.insula &           M<F &  1e-13 &  0.24 &  0.61 &    0.62 &  2e-10 \\
40  &           lh.parsopercularis &                    lh.insula &           M<F &  1e-13 &  0.24 &  0.62 &    0.62 &  3e-10 \\
41  &           rh.superiorfrontal &               rh.paracentral &           M<F &  2e-13 &  0.24 &  0.62 &    0.62 &  3e-10 \\
42  &                Right-Putamen &               Right-Pallidum &           M<F &  2e-13 &  0.24 &  0.62 &    0.62 &  4e-10 \\
43  &                Right-Caudate &      lh.rostralmiddlefrontal &           M>F &  2e-13 &  0.24 &  0.60 &    0.62 &  4e-10 \\
44  &               rh.postcentral &                  rh.bankssts &           M<F &  3e-13 &  0.24 &  0.60 &    0.62 &  5e-10 \\
45  &      lh.rostralmiddlefrontal &           lh.superiorfrontal &           M>F &  3e-13 &  0.24 &  0.62 &    0.62 &  5e-10 \\
46  &  rh.rostralanteriorcingulate &   lh.caudalanteriorcingulate &           M>F &  3e-13 &  0.23 &  0.63 &    0.62 &  6e-10 \\
47  &                   lh.lingual &        lh.transversetemporal &           M<F &  7e-13 &  0.23 &  0.61 &    0.62 &  1e-09 \\
48  &             rh.pericalcarine &            Right-Hippocampus &           M<F &  1e-12 &  0.23 &  0.62 &    0.61 &  2e-09 \\
49  &                   rh.lingual &                Right-Caudate &           M<F &  1e-12 &  0.23 &  0.62 &    0.61 &  2e-09 \\
50  &  lh.rostralanteriorcingulate &   lh.caudalanteriorcingulate &           M>F &  2e-12 &  0.23 &  0.59 &    0.61 &  4e-09 \\
51  &       rh.caudalmiddlefrontal &        Right-Thalamus-Proper &           M>F &  2e-12 &  0.23 &  0.62 &    0.61 &  4e-09 \\
52  &           rh.superiorfrontal &  rh.rostralanteriorcingulate &           M>F &  2e-12 &  0.23 &  0.60 &    0.61 &  4e-09 \\
53  &        rh.posteriorcingulate &            Right-Hippocampus &           M<F &  3e-12 &  0.22 &  0.61 &    0.61 &  7e-09 \\
54  &        Right-Thalamus-Proper &         Right-Accumbens-area &           M>F &  4e-12 &  0.22 &  0.62 &    0.61 &  8e-09 \\
55  &                 rh.precuneus &                    rh.insula &           M<F &  5e-12 &  0.22 &  0.61 &    0.61 &  9e-09 \\
56  &        Right-Thalamus-Proper &               Right-Pallidum &           M<F &  5e-12 &  0.22 &  0.60 &    0.61 &  1e-08 \\
57  &       rh.caudalmiddlefrontal &                Right-Caudate &           M>F &  7e-12 &  0.22 &  0.61 &    0.61 &  1e-08 \\
58  &           lh.parsopercularis &                 Left-Putamen &           M<F &  7e-12 &  0.22 &  0.61 &    0.61 &  1e-08 \\
59  &         Left-Thalamus-Proper &                 Left-Putamen &           M<F &  8e-12 &  0.22 &  0.61 &    0.61 &  2e-08 \\
60  &       rh.caudalmiddlefrontal &                Right-Putamen &           M>F &  1e-11 &  0.22 &  0.62 &    0.61 &  2e-08 \\
61  &      lh.rostralmiddlefrontal &       lh.caudalmiddlefrontal &           M>F &  1e-11 &  0.22 &  0.61 &    0.61 &  3e-08 \\
62  &           rh.superiorfrontal &                Right-Putamen &           M>F &  1e-11 &  0.22 &  0.60 &    0.61 &  3e-08 \\
63  &        Right-Thalamus-Proper &         Left-Thalamus-Proper &           M<F &  2e-11 &  0.22 &  0.62 &    0.61 &  3e-08 \\
64  &                    lh.cuneus &         Left-Thalamus-Proper &           M<F &  2e-11 &  0.22 &  0.61 &    0.61 &  4e-08 \\
65  &          lh.lateraloccipital &                 Left-Caudate &           M<F &  2e-11 &  0.22 &  0.61 &    0.61 &  4e-08 \\
66  &        rh.posteriorcingulate &        Right-Thalamus-Proper &           M<F &  2e-11 &  0.22 &  0.60 &    0.61 &  4e-08 \\
67  &                Right-Caudate &         Left-Thalamus-Proper &           M<F &  2e-11 &  0.22 &  0.59 &    0.61 &  5e-08 \\
68  &                    lh.insula &                Left-Amygdala &           M>F &  3e-11 &  0.22 &  0.61 &    0.61 &  5e-08 \\
69  &           rh.superiorfrontal &        rh.posteriorcingulate &           M>F &  4e-11 &  0.21 &  0.61 &    0.61 &  8e-08 \\
70  &            Right-Hippocampus &        lh.posteriorcingulate &           M<F &  4e-11 &  0.21 &  0.59 &    0.61 &  9e-08 \\
71  &   rh.caudalanteriorcingulate &  lh.rostralanteriorcingulate &           M>F &  5e-11 &  0.21 &  0.62 &    0.61 &  9e-08 \\
72  &          lh.parstriangularis &             lh.supramarginal &           M<F &  5e-11 &  0.21 &  0.59 &    0.61 &  1e-07 \\
73  &           lh.superiorfrontal &       lh.caudalmiddlefrontal &           M>F &  5e-11 &  0.21 &  0.61 &    0.61 &  1e-07 \\
74  &        rh.posteriorcingulate &      lh.rostralmiddlefrontal &           M>F &  6e-11 &  0.21 &  0.61 &    0.61 &  1e-07 \\
75  &           lh.superiorfrontal &  lh.rostralanteriorcingulate &           M>F &  7e-11 &  0.21 &  0.58 &    0.61 &  1e-07 \\
76  &          rh.parstriangularis &          rh.superiorparietal &           M<F &  8e-11 &  0.21 &  0.58 &    0.61 &  1e-07 \\
77  &  rh.rostralanteriorcingulate &                Right-Caudate &           M>F &  8e-11 &  0.21 &  0.59 &    0.61 &  2e-07 \\
78  &          rh.isthmuscingulate &               Right-Pallidum &           M<F &  8e-11 &  0.21 &  0.59 &    0.61 &  2e-07 \\
79  &             lh.pericalcarine &         Left-Thalamus-Proper &           M<F &  1e-10 &  0.21 &  0.61 &    0.61 &  2e-07 \\
80  &       lh.caudalmiddlefrontal &  lh.rostralanteriorcingulate &           M<F &  1e-10 &  0.21 &  0.59 &    0.61 &  2e-07 \\
81  &           rh.parsopercularis &               rh.postcentral &           M<F &  1e-10 &  0.21 &  0.62 &    0.61 &  2e-07 \\
82  &      lh.rostralmiddlefrontal &        lh.posteriorcingulate &           M>F &  1e-10 &  0.21 &  0.61 &    0.60 &  2e-07 \\
83  &                   lh.lingual &                 Left-Putamen &           M<F &  1e-10 &  0.21 &  0.61 &    0.60 &  2e-07 \\
84  &                Right-Caudate &                lh.precentral &           M<F &  1e-10 &  0.21 &  0.59 &    0.60 &  3e-07 \\
85  &           lh.superiorfrontal &                 Left-Caudate &           M>F &  2e-10 &  0.21 &  0.60 &    0.60 &  3e-07 \\
86  &                   lh.lingual &          lh.superiortemporal &           M<F &  2e-10 &  0.21 &  0.59 &    0.60 &  4e-07 \\
87  &          rh.isthmuscingulate &        Right-Thalamus-Proper &           M<F &  2e-10 &  0.21 &  0.60 &    0.60 &  4e-07 \\
88  &          lh.parstriangularis &               lh.postcentral &           M<F &  2e-10 &  0.21 &  0.59 &    0.60 &  4e-07 \\
89  &             lh.pericalcarine &                    lh.insula &           M<F &  2e-10 &  0.21 &  0.60 &    0.60 &  4e-07 \\
90  &        rh.posteriorcingulate &          rh.isthmuscingulate &           M<F &  3e-10 &  0.21 &  0.59 &    0.60 &  5e-07 \\
91  &                lh.precentral &          lh.superiortemporal &           M<F &  3e-10 &  0.21 &  0.58 &    0.60 &  6e-07 \\
92  &                    rh.cuneus &                  rh.bankssts &           M<F &  4e-10 &  0.20 &  0.60 &    0.60 &  7e-07 \\
93  &             Left-Hippocampus &                Left-Amygdala &           M>F &  4e-10 &  0.20 &  0.60 &    0.60 &  7e-07 \\
94  &                   rh.lingual &                Right-Putamen &           M<F &  4e-10 &  0.20 &  0.61 &    0.60 &  7e-07 \\
95  &          lh.lateraloccipital &        lh.transversetemporal &           M<F &  4e-10 &  0.20 &  0.60 &    0.60 &  7e-07 \\
96  &       lh.caudalmiddlefrontal &         Left-Thalamus-Proper &           M>F &  4e-10 &  0.20 &  0.61 &    0.60 &  8e-07 \\
97  &           rh.parsopercularis &                 rh.precuneus &           M<F &  5e-10 &  0.20 &  0.61 &    0.60 &  1e-06 \\
98  &                 rh.precuneus &               Right-Pallidum &           M<F &  6e-10 &  0.20 &  0.61 &    0.60 &  1e-06 \\
99  &          rh.parstriangularis &          rh.inferiorparietal &           M<F &  6e-10 &  0.20 &  0.58 &    0.60 &  1e-06 \\
100 &   rh.caudalanteriorcingulate &           lh.superiorfrontal &           M>F &  6e-10 &  0.20 &  0.59 &    0.60 &  1e-06 \\
101 &                 rh.precuneus &        Right-Thalamus-Proper &           M<F &  7e-10 &  0.20 &  0.61 &    0.60 &  1e-06 \\
102 &                rh.precentral &                  rh.bankssts &           M<F &  8e-10 &  0.20 &  0.58 &    0.60 &  2e-06 \\
103 &               Right-Pallidum &               Right-Amygdala &           M>F &  1e-09 &  0.20 &  0.59 &    0.60 &  2e-06 \\
104 &        Right-Thalamus-Proper &          lh.superiorparietal &           M<F &  1e-09 &  0.20 &  0.60 &    0.60 &  2e-06 \\
105 &          lh.parstriangularis &          lh.superiorparietal &           M<F &  1e-09 &  0.20 &  0.58 &    0.60 &  3e-06 \\
106 &           lh.superiorfrontal &   lh.caudalanteriorcingulate &           M>F &  2e-09 &  0.20 &  0.60 &    0.60 &  3e-06 \\
107 &                   rh.lingual &          rh.superiortemporal &           M<F &  2e-09 &  0.20 &  0.59 &    0.60 &  3e-06 \\
108 &                    rh.insula &            Right-Hippocampus &           M<F &  2e-09 &  0.20 &  0.59 &    0.60 &  5e-06 \\
109 &           lh.parsopercularis &                 lh.precuneus &           M<F &  2e-09 &  0.20 &  0.60 &    0.60 &  5e-06 \\
110 &      rh.rostralmiddlefrontal &          rh.inferiorparietal &           M<F &  3e-09 &  0.20 &  0.59 &    0.60 &  5e-06 \\
111 &          lh.lateraloccipital &                 Left-Putamen &           M<F &  3e-09 &  0.20 &  0.59 &    0.60 &  5e-06 \\
112 &        Right-Thalamus-Proper &                Right-Caudate &           M<F &  3e-09 &  0.19 &  0.60 &    0.60 &  6e-06 \\
113 &             rh.supramarginal &                   rh.lingual &           M<F &  4e-09 &  0.19 &  0.59 &    0.60 &  7e-06 \\
114 &        Right-Thalamus-Proper &        lh.posteriorcingulate &           M<F &  5e-09 &  0.19 &  0.60 &    0.60 &  1e-05 \\
115 &             rh.pericalcarine &                    rh.insula &           M<F &  5e-09 &  0.19 &  0.58 &    0.60 &  1e-05 \\
116 &                 rh.precuneus &                Right-Caudate &           M<F &  9e-09 &  0.19 &  0.60 &    0.59 &  2e-05 \\
117 &                   rh.lingual &        rh.transversetemporal &           M<F &  1e-08 &  0.19 &  0.58 &    0.59 &  2e-05 \\
118 &                  lh.fusiform &                 Left-Caudate &           M<F &  1e-08 &  0.19 &  0.59 &    0.59 &  2e-05 \\
119 &        Right-Thalamus-Proper &                 Left-Putamen &           M<F &  1e-08 &  0.19 &  0.59 &    0.59 &  2e-05 \\
120 &        lh.posteriorcingulate &         Left-Thalamus-Proper &           M<F &  1e-08 &  0.19 &  0.60 &    0.59 &  2e-05 \\
\bottomrule

\end{longtable}

\section*{The fiber length (fl) weight function}

\begin{figure}[H]
	\centering
	\includegraphics[width=16cm]{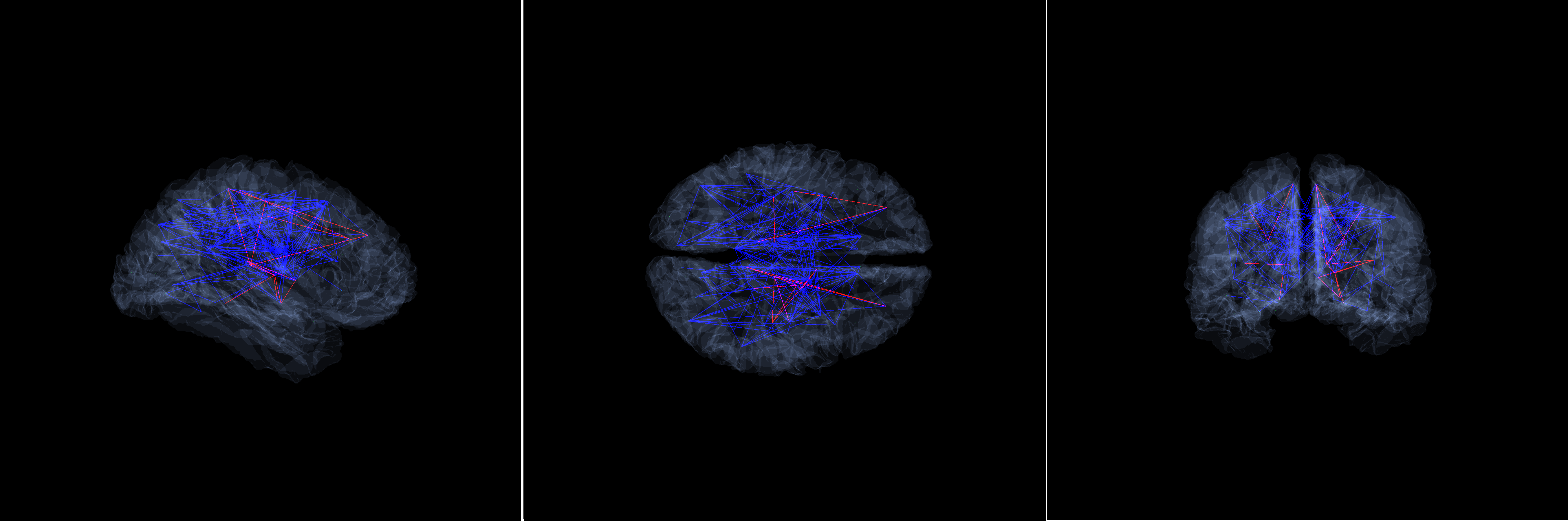}
\end{figure}
\noindent {\bf Figure S4.} {\sl 207 sex implicator edges with the fiber length (fl) weight function with FWER $< 1.95 \times 10^{-5}$. Red color denotes edges where the higher fl weight implies the female sex, while blue edges correspond to those where higher fl weight implies the male sex (cf. Table S5). Three views correspond (from left to right) to sagittal, horizontal, and coronal positions. One may observe that most edges on the figure are blue, since -- statistically -- males have larger brain volumes than females. Most interestingly, there are several red edges on the figure, which are significantly longer in women than in men, and from their length, one can infer the sex of the subject with more than 60\% correctness (e.g., between the Left-Putamen \& Left-Hippocampus, or between right parsopercularis \&  right precentral areas). An animation of these images is available at \url{https://youtu.be/KVzP5OpxU2E}}. 

\section*{Table S5: Sex implicator edges with the fiber length (fl) weight function}

207 sex implicator edges with the fiber length (fl) weight function with FWER $< 1.95 \times 10^{-5}$. Index tells the ranking, \textit{Vertex 1} and \textit{Vertex 2} are the corresponding brain areas, \textit{M>F} indicates an edge with weight function significantly higher in males, while \textit{M<F} indicates an edge with weight function significantly higher in females. The \textit{p\_ks} and \textit{D\_ks} mean the Kolmogorov-Smirnov p-value and statistics for an edge. \textit{ACC} denotes the accuracy of the best separation, \textit{ACC\_ks} denotes the $\frac{1}{2} (1+ \textit{D\_ks})$ KS estimation for best separation accuracy. \textit{FWER} in the $k$th row shows the FWER upper bound for the first $k$ edges. In the statistical analysis only the edges were considered (i.e., where the fl weight function was positive).

{\small
\begin{longtable}{rllclrrrl}
\toprule
{} &              Vertex 1 &              Vertex 2 & Male ? Female &   p\_ks &  D\_ks &   ACC &  ACC\_ks &   FWER \\
\midrule
1   &          lh.superiorfrontal &  lh.caudalanteriorcingulate &           M>F &  2e-42 &  0.42 &  0.71 &    0.71 &  3e-39 \\
2   &          rh.superiorfrontal &  rh.caudalanteriorcingulate &           M>F &  6e-38 &  0.40 &  0.70 &    0.70 &  1e-34 \\
3   &      lh.caudalmiddlefrontal &  lh.caudalanteriorcingulate &           M>F &  1e-31 &  0.36 &  0.68 &    0.68 &  2e-28 \\
4   &          lh.superiorfrontal &               Left-Pallidum &           M>F &  2e-30 &  0.37 &  0.69 &    0.68 &  3e-27 \\
5   &      lh.caudalmiddlefrontal &       lh.posteriorcingulate &           M>F &  7e-30 &  0.35 &  0.68 &    0.68 &  1e-26 \\
6   &      rh.caudalmiddlefrontal &       rh.posteriorcingulate &           M>F &  1e-29 &  0.35 &  0.69 &    0.68 &  2e-26 \\
7   &          rh.superiorfrontal &              Right-Pallidum &           M>F &  2e-27 &  0.35 &  0.68 &    0.68 &  4e-24 \\
\bottomrule

\end{longtable}

\bigskip

\end{document}